\documentclass[a4paper,twocolumn,showpacs,prl]{revtex4}

\usepackage{graphicx}
\usepackage{dcolumn}
\usepackage{bm}
\usepackage{amsmath}
\usepackage{amssymb}
\usepackage{graphicx}
\usepackage{amsfonts}
\usepackage{txfonts}

\begin{document}

\preprint{APS/123-QED}
\title{Direct detection of the relative strength of Rashba and Dresselhaus
spin-orbit interaction: Utilizing the SU(2) symmetry}
\author{Jun Li}
\author{Kai Chang}
\email{kchang@red.semi.ac.cn}
\affiliation{SKLSM, Institute of Semiconductors, Chinese Academy of Sciences, P. O. Box
912, Beijing 100083, China}
\date{\today}

\begin{abstract}

We propose a simple method to detect the relative strength of Rashba
and Dresselhaus spin-obit interactions in quantum wells (QWs)
without relying on the directional-dependent physical quantities.
This method utilize the asymmetry of critical gate voltages that
leading to the remarkable signals of SU(2) symmetry, which happens
to reflect the intrinsic structure inversion asymmetry of the QW. We
support our proposal by the numerical calculation of in-plane
relaxation times based on the self-consistent eight-band Kane model.
We find that the two different critical gate voltages leading to the
maximum spin relaxation times [one effect of the SU(2) symmetry] can
simply determine the ratio of the coefficients of Rashba and
Dresselhaus terms. Our proposal can also be generalized to extract
the relative strengths of the spin-orbit interactions in quantum
wire and quantum dot structures.
\end{abstract}

 \maketitle


The spin-obit interaction (SOI), which is a manifestation of the
relativistic effect, transforms the electric fields into
momentum-dependent effective magnetic fields, coupling the electron
spin with electron orbital motion. The SOI provides us an efficient
way to control the electron spin with electric fields instead of
magnetic fields\cite{Nitta, SHE}, therefore plays an important role
in realizing all-electrical controlled spintronic devices. According
to the different origins of SOI in semiconductor quantum structures,
the SOI has been distinguished by the Rashba SOI (RSOI) arising from
structure inversion asymmetry (SIA)\cite{RSOI} and the Dresselhaus
SOI (DSOI) caused by bulk inversion asymmetry (BIA),
respectively.\cite{DSOI} These two types of SOI, yielding different
effective SO magnetic field, leads to different behaviors of
spin-transport properties and spin relaxation. Naturally, the ratio
of Rashba and Dresselhaus coefficients (RD ratio) becomes a key
parameter for understanding the spin-related phenomena and designing
the future spintronic devices. Previously, the RD ratio can be
determined by mapping the $\boldsymbol{k}$-dependent spin photocurrent,\cite%
{Ganichev, Ganichev2} in-plane spin-relaxation
time,\cite{AverkievExp} the spin precession about the effective
spin-obit magnetic fields\cite{Meier} into the components coming
from DSOI and RSOI, or utilizing the anisotropic conductance of
quantum wires in the presence of in-plane magnetic
fields.\cite{Scheid} Therefore the above methods require exquisite
measurement with respect of the crystallographic axis. Although
these methods have been successfully used to study the relative
strength of the Rashba and Dresselhaus SOIs in two-dimensional
quantum well and heterostructures, the obtained RD ratio still holds
certain ambiguity as pointed out by the authors\cite{Ganichev2,
AverkievExp, Scheid}. Hence, finding a simple and accurate method to
determine the RD ratio in all sorts of systems remains a challenging
but important task.

\begin{figure}[h]
\includegraphics[width=\columnwidth] {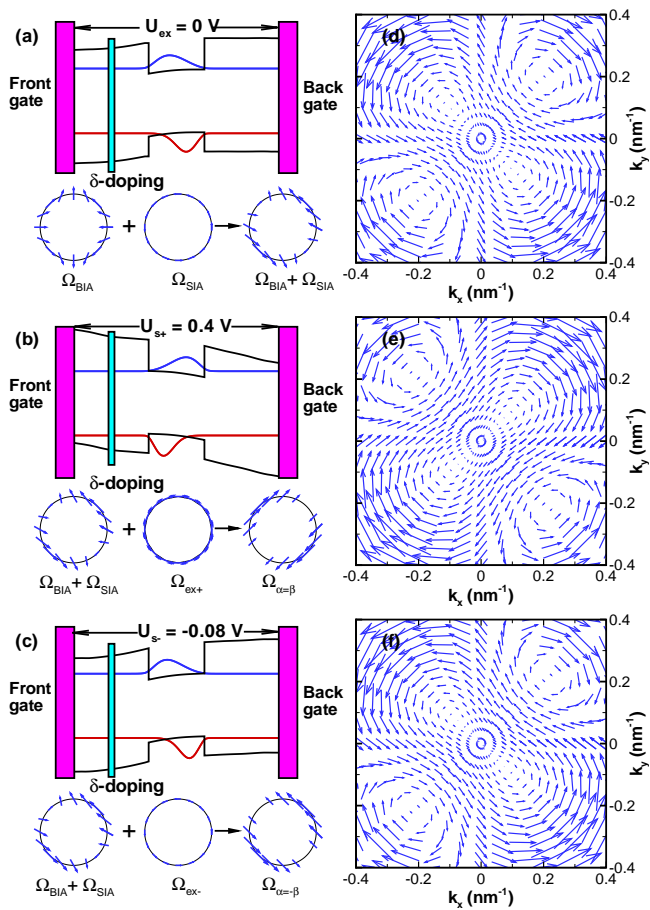}
\caption{(Color online) The calculated band profiles of an asymmetrically n-doped Al$_{0.3}$%
Ga$_{0.7}$As/GaAs/Al$_{0.3}$Ga$_{0.7}$As QW and the electron (hole)
probability distribution for different gate voltage bias (a)
$U_{ex}=0$ V (b) $U_{s+}=0.4$ V (c) $U_{s-}=-0.08$ V. The sketches
under each band profiles show schematically the effective spin-obit
magnetic field. The panels (d) (e) (f) are the effective spin-obit
magnetic field calculated by self-consistent eight-band Kane model
corresponding to (a) (b) (c), respectively. The doping concentration
is fixed at $N_{D}=4.0\times 10^{11}$ cm$^{-2}$.} \label{fig:fig1}
\end{figure}

In this Letter, we propose a direct method that can separate the
RSOI from DSOI and determine the RD ratio in an asymmetric
[001]-oriented zincblende quantum well (QW). Applying a gate voltage
cross the QW to tune the total RSOI in this structure (see Fig.
\ref{fig:fig1}), we can find two different
magnitudes of critical gate voltages to restore the exact SU(2) symmetry\cite%
{Bernevig} by strengthening or canceling intrinsic RSOI existing in
this QW. The difference between these two critical gate voltages
extract the exact information of intrinsic structure inversion
asymmetry of this QW, with bulk inversion asymmetry separating
apart. Therefore the two critical gate voltages can be used to
determine the RD ratio in this asymmetric QW. Compare with the
previous works, this proposal does not rely on measurements along
specific directions\cite{Explain} and is robust against all the
effects that cannot change the structure inversion asymmetry such as the isotropic impurity scattering.
In addition, this proposal offers a general scheme that does not
dependent on a specific experimental technology and the
dimensionality of the experimental sample, e.g., quantum wires and
dots. A series of remarkable physical effects in quantum wells, wires and dots\cite%
{Averkiev, Bernevig,PSHexp, WAL1, WAL2, SdH1,SdH2, Qwire1, Qwire2, Qdot}
led by the SU(2) symmetry can be used to measure the critical gate
voltages, consequently the RD ratio in these quantum structures.

Firstly we give a picture of our proposal based on the single-band model of
two-dimensional electron gas (2DEG) with two types of SOI:

\begin{equation}
H=\frac{\hbar ^{2}k^{2}}{2m}+H_{R}+H_{D}  \label{eq:H2band}
\end{equation}%
where $H_{R}=\alpha \left( k_{y}\sigma _{x}-k_{x}\sigma _{y}\right) $ is the
Rashba spin-obit interaction term, $H_{D}=\beta \left( k_{y}\sigma
_{y}-k_{x}\sigma _{x}\right) +\gamma \left( k_{y}^{2}k_{x}\sigma
_{x}-k_{x}^{2}k_{y}\sigma _{y}\right) $ is the Dresselhaus spin-obit
interaction term, and $\boldsymbol{k}=\left( k_{x},k_{y}\right) $ is
in-plane wave vector. Here, $\alpha $ is the linear Rashba coefficient, $%
\beta $ and $\gamma $ are the linear and cubic Dresselhaus coefficients,
respectively. The Rashba coefficient $\alpha $ can be tuned easily by
changing the structure inversion asymmetry, for instance, by gate voltage
applied perpendicular to the QW plane.\cite{Nitta} While the Dresselhaus
coefficients $\beta $ and $\gamma $ can be adjusted by changing the
thickness of quantum wells. If we adopt the infinite high barrier model, $%
\beta \approx $ $\gamma \left\langle k_{z}^{2}\right\rangle =\gamma \left(
\frac{\pi }{w}\right) ^{2}$.

The interplay between the RSOI and DSOI would lead to the anisotropy of
optical and transport properties, since the DSOI depends sensitively on the
crystallographic orientations, while the RSOI shows an isotropic behavior.
If we tune gate voltage to satisfy $\alpha =\pm \beta $ (neglecting the
cubic Dresselhaus term), the Hamiltonian of 2DEG show the exact SU(2)
symmetry.\cite{Bernevig} The exact SU(2) symmetry is a very unique property
of quantum systems that the RSOI and DSOI happen to cancel each other for $%
\boldsymbol{k}$ along $[110]$ or $[1\overline{1}0]$ and is revealed to be
robust against spin-independent disorder interactions.\cite{Bernevig} As a
consequence, the exact SU(2) symmetry would lead to a series of remarkable
physical effects. For example, there should be a maximum spin life-time for
electron spins align along $[1\overline{1}0]$ or $[110]$ direction,\cite%
{Averkiev} a persistent spin helix could exist in the sample,\cite%
{Bernevig,PSHexp} the diminishing of the weak
antilocalization\cite{WAL1, WAL2} and the beating pattern of SdH
oscillation.\cite{SdH1,SdH2} Besides it is worth noticing that in
quasi-one-dimensional quantum wire and zero-dimensional quantum dot,
the SU(2) symmetry could also induce strong physical effects. Such as the conductance of a quantum wire shows strong anisotropy\cite{Qwire1,
Qwire2} and the spin relaxation curve of a quantum dot shows a cusplike structure\cite{Qdot}. All these physical effects can be used to determine the
critical gate voltages that restore the SU(2) symmetry.

Notice that the SU(2) symmetry could be achieved by applying both positive
and negative electric fields, each satisfying $\alpha =\pm \beta $. The
total Rashba coefficient of an asymmetric QW with gate voltage tuning the
SIA can be viewed as a superposition of two parts $\alpha =\alpha
_{0}+\alpha _{ex}$. The first part $\alpha _{0}$ comes from the intrinsic
SIA of the sample, e.g., the asymmetric doping or band profile. The second
part $\alpha _{ex}$ is introduced by the external electric field of the gate
voltage. By sweeping the gate voltage, one can find two values $U_{s\pm }$
to meet the SU(2) symmetry conditions: $\alpha _{0}+\alpha _{s\pm }=\pm
\beta $. Here we label the external Rashba coefficients that lead to the
SU(2) symmetry conditions with $\alpha _{s\pm }$. If we neglect the
difference of the dielectric constant between the well and barrier
materials, we can simply assume $\alpha _{s\pm }$ to be proportional to $%
U_{s\pm }$. From the requirement restoring the SU(2) symmetry, one can find

\begin{equation}
\frac{\alpha _{0}}{\beta }=\frac{\alpha _{s+}+\alpha _{s-}}{\alpha
_{s-}-\alpha _{s+}}=\frac{U_{s+}+U_{s-}}{U_{s-}-U_{s+}}.  \label{eq:Ratio}
\end{equation}

Eq. (\ref{eq:Ratio}) demonstrate that, if there is no intrinsic SIA
in the sample, i.e., $\alpha _{0}=0$, we should expect that
$U_{s+}=-U_{s-}$; while if $\alpha _{0}\neq 0$, we get $U_{s+}\neq
-U_{s-}$. From a symmetry consideration, this conclusion is easy to
understand because there would be no difference between $\left\vert
U_{s\pm }\right\vert $ unless the $[001]$ and $[00\overline{1}]$
directions of the QW are asymmetric. This consideration guarantee
our proposal to be robust against all the effects that does not
change the symmetry of $[001]$ and $[00\overline{1}]$ directions,
such as the isotropic impurity scattering. This conclusion can be
also supported by an eight-band self-consistent calculation, as
shown in
Fig. \ref{fig:fig1}. In an asymmetrically doped Al$_{0.3}$Ga$_{0.7}$%
As/GaAs/Al$_{0.3}$Ga$_{0.7}$As QW, the results show that the critical
voltages to satisfy $\alpha =\pm \beta $ are $0.4$ V and $-0.08$ V
respectively [see the panels (b) and (c)]. From the compositions of total
effective spin-orbit magnetic field (the panels under the band profile), one
can see clearly that the difference between $\left\vert U_{s\pm }\right\vert
$ comes from the intrinsic SIA. In Fig. \ref{fig:fig1} (d)-(f), we show the
effective spin-orbit magnetic field calculated by eight-band self-consistent
calculation, which already takes the cubic Dresselhaus terms into account.
Although the existence of cubic Dresselhaus terms might cause a different
configuration rather than the exact SU(2) symmetry [see the panels (e) and
(f)], the panel (e) still shows the mirror reflection symmetry with the
panel (f), indicating that the total Rashba coefficient of these two panels
are of the same magnitude (but with opposite signs). The intrinsic SIA still
request two asymmetrical critical voltages to achieve the total ROI in
panels (e) and (f). So the asymmetrical critical voltages always reflects
the intrinsic SIA of QW, even taking account of the cubic Dresselhaus terms.

\begin{figure}[th]
\includegraphics[width=\columnwidth] {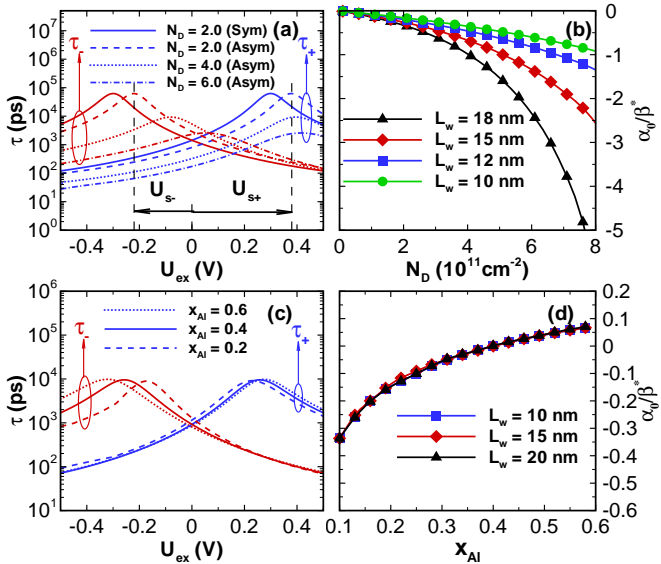}
\caption{(Color online) (a) Calculated in-plane spin relaxation
times $\protect\tau _{\pm }$
as a function of gate voltage in n-doped Al$_{0.3}$Ga$_{0.7}$As/GaAs/Al$%
_{0.3}$Ga$_{0.7}$As QW with different doping conditions: The solid
lines denote the QW doped symmetrically; The dashed, dotted and
dashdotted lines denote the QWs doped asymmetrically with different
doping concentrations (the unit is 10$^{11}\,$cm$^{-2}$) (c)
$\protect\tau _{\pm }$ as a function of gate voltage in
Al$_{x}$Ga$_{1-x}$As/GaAs/Al$_{0.3}$Ga$_{0.7}$As QW with
different Al composition $x$. (b) and (d) the RD ratios $\protect\alpha _{0}/%
\protect\beta ^{\ast }$ as a function of the doping concentration and Al
composition, respectively, for different thicknesses of QWs. }
\label{fig:fig2}
\end{figure}

Next, we will take the in-plane D'yakonov-Perel' (DP) spin relaxation times%
\cite{DP} as an example to demonstrate the validity of our proposal. This
calculation is based on a self-consistent eight-band Kane model.\cite%
{InSbSRT} The band parameters can be found in Ref. \onlinecite{Parameters},
and the BIA Kane parameter $B_{0}$ are obtained from 14-band effective mass
model in Ref. \onlinecite{SOIWinkler}. In Fig. \ref{fig:fig2} (a) and (c) we
exhibit the calculated spin relaxation times for electron spin along $[110]$
or $[1\overline{1}0]$ (denoted by $\tau _{+}$, $\tau _{-}$ respectively) as
a function of gate voltage in 15 nm n-doped GaAs/AlGaAs QWs with different
doping conditions and Al compositions of barrier (i.e., asymmetrical barrier
heights), respectively. In order to understand the numerical results of spin
relaxation times, we introduce the analytical results of DP spin relaxation
times at T = 0 K\cite{Averkiev}
\begin{equation}
\frac{1}{\tau _{\pm }}=\frac{2\tau _{1}}{\hbar ^{2}}\left[ \left( \alpha \mp
\beta \,\right) ^{2}k_{F}^{2}-\frac{1}{2}\gamma \left( \beta \mp \alpha
\right) k_{F}^{4}+\frac{1+\tau _{3}/\tau _{1}}{16}\gamma ^{2}k_{F}^{6}\right]
,  \label{eq:Taupm}
\end{equation}%
where $k_{F}$ is the Fermi wave vector of electron, and we simply take a
typical momentum scattering time $\tau _{1}=\tau _{3}=0.1$ ps in this paper.
By sweeping the gate voltage, the total Rashba coefficient $\alpha $ change
linearly. So one can find the maximum spin relaxation time $\tau _{\pm
}^{\max }=8\hbar ^{2}/\left( k_{F}^{6}\gamma ^{2}\tau _{3}\right) $ when $%
\alpha =\pm \left( \beta -\frac{1}{4}k_{F}^{2}\gamma \right) $,
corresponding to the peaks of $\tau _{\pm }$ in Fig. \ref{fig:fig2} (a). The
two peaks are symmetric with respect to the zero voltage ($\left\vert
U_{s+}\right\vert =\left\vert U_{s-}\right\vert $) for a symmetrical QW [see
the solid lines] but asymmetric ($\left\vert U_{s+}\right\vert \neq
\left\vert U_{s-}\right\vert $) [see the dashed, dotted and dashdotted
lines] for asymmetrical QWs. Different from the $k$-linear SOI model which
neglects the cubic Dresselhaus term, $\tau _{\pm }$ would not go infinite
and the maximum values do not show at $\alpha =\pm \beta $. Similar to the
analysis on the spin-galvanic effect by Ganichev et al.\cite{Ganichev}, we
can take the $\beta ^{\ast }=\beta -\frac{1}{4}k_{F}^{2}\gamma $ as the
renormalized Dresselhaus coefficient for $k$-linear SOI model. Therefore we
actually get $\alpha _{0}/\beta ^{\ast }$ from Eq. (\ref{eq:Ratio}) rather
than $\alpha _{0}/\beta $ and the difference between $\beta \,$and $\beta
^{\ast }$ comes from the contribution of cubic Dresselhaus term. In Figs. %
\ref{fig:fig2} (b) and (d) we display $\alpha _{0}/\beta ^{\ast }$
determined from the critical gate voltage that lead to the maximum in-plane
spin relaxation time. As expected, $\alpha _{0}/\beta ^{\ast }\,$increase
with increasing the asymmetrical doping concentration or the composition
difference between the left and right barriers. For the QWs with different
doping concentrations, $\alpha _{0}/\beta ^{\ast }\,$also increase with
increasing the well width, since $\beta ^{\ast }\sim \beta \approx \gamma
\left( \frac{\pi }{w}\right) ^{2}$. For the QWs with different compositions
of barriers, $\alpha _{0}/\beta ^{\ast }$ turns to be not sensitive to the
change of well width, because $\alpha _{0}$ are very small in these cases.

\begin{figure}[t]
\includegraphics[width=\columnwidth] {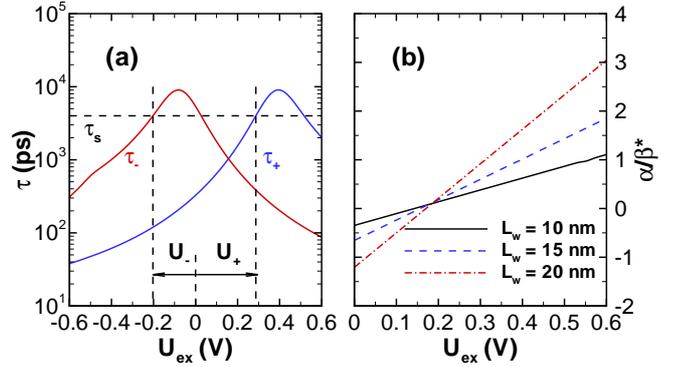}
\caption{(Color Online) (a) The in-plane spin relaxation time
$\protect\tau _{\pm }$ as a
function of gate voltage in 15 nm asymmetrically n-doped Al$_{0.3}$Ga$_{0.7}$%
As/GaAs/Al$_{0.3}$Ga$_{0.7}$As QW ($\,N_{D}\,=4\times 10^{11}$ cm$^{-2}$)
(b) $\protect\alpha /\protect\beta ^{\ast }$ determined by Eq. (\protect\ref%
{eq:totalRD1}) as a function of gate voltage for Al$_{0.3}$Ga$_{0.7}$%
As/GaAs/Al$_{0.3}$Ga$_{0.7}$As QWs with different well widths.}
\label{fig:fig3}
\end{figure}

In addition to the intrinsic Rashba coefficient $\alpha _{0}$, it is also
possible to determine the relative strength of total Rashba coefficient $%
\alpha $ and Dresselhaus coefficient $\beta $ in the QW through the in-plane
spin relaxation times. From Eq. (\ref{eq:Taupm}) we can find the condition
for the in-plane spin relaxation times $\tau _{\pm }$ to reach a same value $%
\tau _{s}$ satisfying $\alpha _{0}+\alpha _{\pm }=\pm \alpha $. As shown in
Fig. \ref{fig:fig3} (a), for a $\tau _{s}<\tau _{\pm }^{\max }$, we can find
the gate voltages $U_{\pm }$ corresponding to $\tau _{\pm }=\tau _{s}$.
Usually, there are two different gate voltages for each $\tau _{+}$ and $%
\tau _{-}$. So we must limit that if $\left\vert U_{+}\right\vert
>\left\vert U_{s+}\right\vert $ ($\left\vert U_{+}\right\vert <\left\vert
U_{s+}\right\vert $), we choose $\left\vert U_{-}\right\vert >\left\vert
U_{s-}\right\vert $ ($\left\vert U_{-}\right\vert <\left\vert
U_{s-}\right\vert $). Combine these conditions and Eq. (\ref{eq:Ratio}), we
can get
\begin{equation}
\frac{\alpha }{\beta ^{\ast }}=\frac{\left( U_{s+}+U_{s-}\right) \left(
U_{-}-U_{+}\right) }{\left( U_{s-}-U_{s+}\right) \left( U_{+}+U_{-}\right) }.
\label{eq:totalRD1}
\end{equation}%
If the QW is inversion symmetric, we should have
\begin{equation}
\frac{\alpha }{\beta ^{\ast }}=\frac{U_{-}-U_{+}}{U_{s+}-U_{s-}}=\frac{%
U_{\pm }}{U_{s\pm }}.  \label{eq:totalRD2}
\end{equation}%
In Fig. \ref{fig:fig3} (b) we plot $\alpha /\beta ^{\ast }$ as a function of
gate voltage in asymmetrically n-doped GaAs/AlGaAs QWs with different well
widths. The figure shows $\alpha /\beta ^{\ast }$ increase almost linearly
with gate voltage, which is consistent with the analytical results of Rashba
coefficient in the previous work\cite{RSS}. The slope of $\alpha /\beta
^{\ast }$ for narrow wells are smaller than that of thick wells because $%
\beta ^{\ast }$ is larger in narrow wells.

\begin{figure}[t]
\includegraphics[width=\columnwidth] {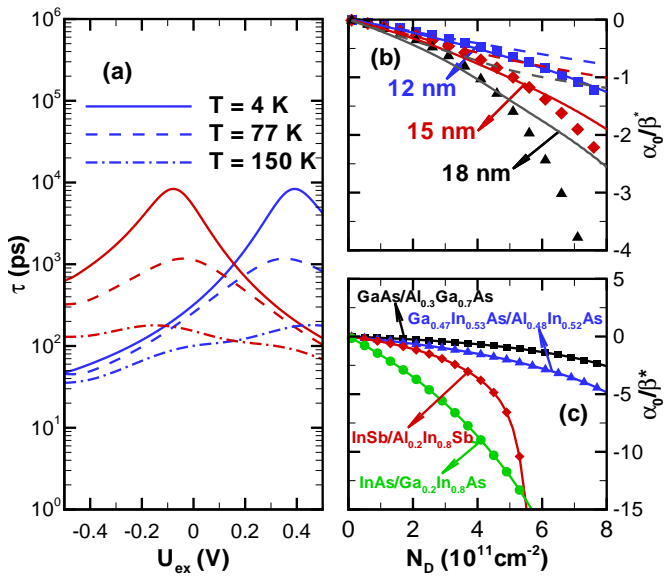}
\caption{(Color online) (a) Calculated in-plane spin relaxation time
$\protect\tau _{\pm }$
as a function of gate voltage in an 15 nm asymmetrically doped ($%
\,N_{D}=4\times 10^{11}$ cm$^{-2}$) Al$_{0.3}$Ga$_{0.7}$As/GaAs/Al$_{0.3}$Ga$%
_{0.7}$As QW at different temperatures. (b) The RD ratio $\protect\alpha %
_{0}/\protect\beta ^{\ast }$ as a function of asymmetric doping
concentration in Al$_{0.3}$Ga$_{0.7}$As/GaAs/Al$_{0.3}$Ga$_{0.7}$As QW with
different well widths $L_{w}=12,15,18$ nm. The triangle, diamond and square
dots are obtained from Eq. (\protect\ref{eq:Ratio}). The solid and dashed
lines are $\protect\alpha _{0}/\protect\beta ^{\ast }$\thinspace and $%
\protect\alpha _{0}/\protect\beta $ obtained by fitting parameters directly
from the eight-band spin-splitting, respectively. (c)The RD Ratio $\protect%
\alpha _{0}/\protect\beta ^{\ast }$ obtained from Eq. (\protect\ref{eq:Ratio}%
) as a function of asymmetrical doping concentration $N_{D}$ for different
material of QWs.}
\label{fig:fig4}
\end{figure}

In Fig. \ref{fig:fig4} (a), we investigate the temperature effect on the
in-plane spin relaxation time. As we shown in the figure, the in-plane spin
life time peaks which characters the emergence of SU(2) symmetry would be
gradually smeared out when temperature increases from $T=4$ K to $150$ K or
even higher. That is because of the blurring of the Fermi surface with
increasing temperature.\cite{InSbSRT} We may also associate that many of
other SU(2) symmetry phenomenons may disappear at high temperature due to
the blurring of the Fermi surface. So it is suggested that the SU(2)
symmetry phenomenons should be observed at $T<77$ K. In Fig. \ref{fig:fig4}
(b) we compare the RD ratios $\alpha _{0}/\beta ^{\ast }$ obtained by Eq. (%
\ref{eq:Ratio}) and that obtained by fitting parameters directly from the
spin-splitting. We find that the ratios $\alpha _{0}/\beta ^{\ast }$ agree
well with the results fitted from the spin-splitting at low doping
concentration and are especially good for narrow QWs. For heavily doped QW
with wide wells, $\alpha _{0}/\beta ^{\ast }$ obtained by the spin
relaxation time calculation could deviate from that from the fitting of the
spin-splitting. The reason is when the doping concentration is high, $\alpha
_{0}$ could be influenced by the change of the internal electric field due
to the charge redistribution induced by the external gate voltage. That
makes the $\alpha _{0}$ under the external electric field different from the
direct fitting parameters from the zero-field spin-splitting. However, as we
see in the Fig. \ref{fig:fig4} (b), the charge redistribution effect is very
limited at light doping ($N_{D}<4\times 10^{11}$ cm$^{-2}$) and narrow QWs $%
\left( L_{w}<15\text{ nm}\right) $. $\alpha _{0}/\beta ^{\ast }$ obtained by
the critical gate voltages measurement are more accurate in these cases. In
Fig. \ref{fig:fig4} (b) we also show the difference between $\alpha
_{0}/\beta $ and $\alpha _{0}/\beta ^{\ast }$. This difference comes from
the contribution of the cubic Dresselhaus terms, and is very small (less
than 0.2) when $N_{D}$ is less than $3\times 10^{11}$ cm$^{-2}$, which
demonstrates that the single band model with $k$-linear SOI coefficients are
valid at low doping concentration. In Fig. \ref{fig:fig4} (c) we show the
calculated RD ratio $\alpha _{0}/\beta ^{\ast }$ for different QWs. Though
these BIA Kane parameters $B_{0}$ are still in a big uncertainty today, we
can still see that for the narrow bandgap QW, such as InAs/GaInAs and
InSb/AlInSb, the relative strength of Rashba SOI are much larger than that
of middle bandgap QWs GaAs/AlGaAs, GaInAs/AlInAs. This is because the RSOI\
comes from the interband coupling of conduction and valence bands, which is
much stronger in narrow bandgap materials.

In summary, we proposed a simple and direct method to separate the
intrinsic RSOI from DSOI. The relative strength of RSOI and DSOI\
can be determined by the critical gate voltages that restores the
SU(2) symmetry in 2DEG. The SU(2) symmetry leads to a series of
characteristic physical effects, such as the maximum in-plane spin
relaxation time, the persistent spin helix and so on. Through the
in-plane spin relaxation time calculation based on the
self-consistent eight-band model, we demonstrate our proposal is
valid and can be used to detect the strengths of the SOIs in quantum
wells, wires and dots utilizing the SU(2) symmetry. Our proposal
offers a general scheme that many experimental techniques could be
used to determine this important parameter and facilitate us to
manually control the spin degree of freedom.

\begin{acknowledgments}
This work is supported by the NSFC Grant No. 60525405, 10874175 and National
Basic Research Program of China(973 Program) (2010CB933700)
\end{acknowledgments}


\begin{thebibliography}{99}
\bibitem{SpintronicsRMD} I. \v{Z}uti\'{c}, J. Fabian, and S. Das Sarma, Rev.
Mod. Phys. \textbf{76}, 323 (2004).

\bibitem{Nitta} J. Nitta, T. Akazaki, H. Takayanagi, and T. Enoki, Phys.
Rev. Lett. \textbf{78}, 1335 (1997).

\bibitem{SHE} W. Yang, K. Chang, and S. C. Zhang, Phys. Rev. Lett. \textbf{\
100}, 056602 (2008).

\bibitem{RSOI} E. I. Rashba, Sov. Phys. Solid State \textbf{2}, 1109 (1960).

\bibitem{DSOI} G. Dresselhaus, Phys. Rev. \textbf{100}, 580 (1955).

\bibitem{Ganichev} S. D. Ganichev, et al., Phys. Rev. Lett. \textbf{92}, 256601
(2004).

\bibitem{Ganichev2} S. Giglberger, et al., Phys. Rev. B \textbf{75},
035327 (2007).

\bibitem{AverkievExp} N. S. Averkiev, et al., Phys. Rev. B \textbf{74}, 033305 (2006).

\bibitem{Meier} L. Meier, et al., Nat. Phys. \textbf{3}, 650 (2007).

\bibitem{Scheid} M. Scheid, M. Kohda, Y. Kunihashi, K. Richter, and J.
Nitta, Phys. Rev. Lett. 101, 266401 (2008).

\bibitem{Bernevig} B. A. Bernevig, J. Orenstein, and S.-C. Zhang, Phys. Rev.
Lett. \textbf{97}, 236601 (2006).

\bibitem{Explain} Some effects of the SU(2) symmetry (mostly are the
anisotropic properties) needs to be observed by the directional-dependent
quantities, but the values of critical gate voltages and the RD ratio are
not sensitive to the directional measurement.

\bibitem{Averkiev} N. S. Averkiev and L. E. Golub, Phys. Rev. B \textbf{60},
15582 (1999); N. S. Averkiev, L. E. Golub, and M. Willander, J. Phys.:
Condens. Matter \textbf{14}, R271 (2002).

\bibitem{PSHexp} J. D. Koralek, et al., Nature \textbf{458}, 610 (2009).

\bibitem{WAL1} W. Knap, et al., Phys. Rev. B \textbf{53}, 3912 (1996).

\bibitem{WAL2} J. B. Miller, et al., Phys.
Rev. Lett. \textbf{90}, 076807 (2003).

\bibitem{SdH1} S. A. Tarasenko and N. S. Averkiev, Pis'ma Zh. Eksp. Teor.
Fiz. \textbf{75}, 669 (2002) [JETP Lett. \textbf{75}, 552 (2002)].

\bibitem{SdH2} W. Yang and K. Chang, Phys. Rev. B \textbf{73}, 045303 (2006).

\bibitem{Qwire1} M. Wang, K. Chang, and K. S. Chan, Appl. Phys. Lett.
\textbf{94}, 052108 (2009).

\bibitem{Qwire2} M. Wang, K. Chang, et al., Nanotechnology \textbf{20}, 365202 (2009).

\bibitem{Qdot} D. V. Bulaev and D. Loss, Phys. Rev. B \textbf{71}, 205324 (2005).


\bibitem{DP} D'yakonov, M. I., and V. I. Perel', Fiz. Tverd. Tela \textbf{13}%
, 3581 [Sov. Phys. Solid State \textbf{13}, 3023(1971)].

\bibitem{InSbSRT} J. Li, K. Chang, and F. M. Peeters, Phys. Rev. B \textbf{80%
}, 153307 (2009).

\bibitem{Parameters} I. Vurgaftman, J. R. Meyer, and L. R. Ram-Mohan, J.
Appl. Phys. \textbf{89}, 5815 (2001).

\bibitem{SOIWinkler} R. Winkler, \emph{Spin-Orbit Coupling Effects in
Two-Dimensional Electron and Hole Systems} (Springer-Verlag, Berlin, 2003),
Chap. 6, pp. 74.

\bibitem{RSS} W. Yang and K. Chang, Phys. Rev. B \textbf{73}, 113303 (2006).
\end{thebibliography}
\end{document}